\documentclass[twocolumn]{jpsj3}
\usepackage{bm,amsmath,amssymb,stmaryrd} \usepackage{graphicx}
\usepackage{txfonts}

\renewcommand{\Im}{\mathop{\mathrm{Im}}\nolimits}

\begin{document}

\title{Universal conductance enhancement and reduction of the
  two-orbital Kondo effect}

\author{Haruka Oguchi and Nobuhiko Taniguchi}

\inst{Institute of Physics, University of Tsukuba, Tennodai
  Tsukuba 305-8571, Japan}

\date{\today}

\abst{
  We investigate theoretically the linear and nonlinear conductance
  through a nanostructure with two-fold degenerate single levels,
  corresponding to the transport through nanostructures such as a
  carbon nanotube, or double dot systems with capacitive interaction.
  It is shown that the presence of the interaction asymmetry between
  orbits/dots affects significantly the profile of the linear
  conductance at finite temperature, and, of the nonlinear
  conductance, particularly around half-filling, where the two-particle
  Kondo effect occurs.  Within the range of experimentally feasible
  parameters, the $SU(4)$ universal behavior is suggested,  and
  comparison with relevant experiments is made. 
} 


\kword{carbon nanotube, quantum dot, nonlinear conductance, Kondo effect, Coulomb interaction, Anderson impurity model}


\maketitle%

\section{Introduction}

Electrical transport through nanostructure devices or a quantum dot is
fundamentally influenced by the presence of the Coulomb interaction on
the dot, where along with the Coulomb blockade phenomena, the Kondo
effect, a paradigm of the strongly correlated phenomena, has been
observed~\cite{Goldhaber-Gordon98}.  The great advantage of these
nanostructure systems is the ability to control various relevant
parameters to regulate many-body effect.  Toward future applications of
quantum electron devices, interest in a role of
many-body effect of the transport phenomena has been rekindled and
studies have been made actively.

In the absence of magnetic field, the levels of a quantum dot are spin
degenerate; the Kondo effect gives rise to the conductance enhancement
at low-temperatures in an odd electron number on the dot, as a result
of the interaction between a local magnetic moment of the dot and
conduction electrons.  The spin degree can be replaced by another
degenerate degrees of freedom.  Indeed, two quantum dots with
capacitive interaction can be labeled by an index $i=1,2$, which can
be taken as another realization of the Anderson impurity
model~\cite{Wilhelm01}; the orbital Kondo effect
appears~\cite{Wilhelm00,Wilhelm02,Chan02,McClure07,Hubel07,Hubel08,Pohjola01,Boese02}.
Other notable examples are vertical quantum dots~\cite{Sasaki00}, or
carbon nanotube
dots~\cite{Nygard00,Babic04,Jarillo-Herrero05,Makarovski07}.  All of
these systems accommodate one, two, three, four electrons in the
topmost shell, and the Coulomb blockade peaks show a clear sign of the
fourfold pattern.
When the spin and orbital degeneracies are simultaneously present, the
further entanglement between spin and orbital degrees of freedom
occurs, giving rise to the $SU(4)$ Kondo effect with an Kondo
temperature one order of magnitude higher than the standard case~\cite{Pohjola01,Boese02,Borda03,Choi05,Lim06,Wilhelm00,Wilhelm02,Jarillo-Herrero05,Makarovski07}.

Though most of studies of the systems with doubly degenerate orbitals have been
focusing on odd-number valleys ($N_{d}\approx 1$ or $3$), it
has been recognized recently that a similar Kondo-like enhancement
(two-particle $SU(4)$ Kondo effect) can occur on even-number valley
($N_{d}\approx 2$) as well~\cite{Galpin05,Galpin06} (See
also ref.~\citen{Eto00} for the singlet-triplet Kondo effect.)
However, such enhancement at even-number valleys does not seem
ubiquitous, observed in some experiments but not in others.  The
conductance of carbon nanotube dots fits well with $G \propto
\sin^{2}(\pi N_{d}/4)$ at very low temperature~\cite{Makarovski07},
which shows the Kondo enhancement both at odd and even valleys; but in two
coupled dots, the honeycomb-like structure as a function of
the two gate voltages shows the enhancement at $N_{d}=1$ but not
at $N_{d}=2$.

In this paper, we will examine the quantum transport through a dot
with doubly degenerate orbitals with different interaction strengths
within the orbits and between orbits.  We will show that such
interaction asymmetry, if small, is important to understand the
profile of the linear conductance at finite temperature, and, of the
nonlinear conductance.  Resorting to the universality of the topmost
shell, the $SU(4)$ Anderson model, we analyze how the conductance gets
enhanced or suppressed at finite temperature or by finite source-drain
voltage.
Though the conductance profiles observed in several experiments may
seem nonuniversal at first glance, we will find that we can explain
them systematically by the $SU(4)$ Anderson model universality, once
taking proper account of the presence of the interaction asymmetry between
orbits/dots.

We are concerned particularly with the universal dependence of the
temperature and/or the bias voltage of the conductance.  
Naively, one may expect the singlet-triplet Kondo effect at
half-filling when the interaction asymmetry is present.  On the other
hand, even with the interaction asymmetry, the renormalization group
(RG) flow is known to drive the system toward the $SU(4)$-symmetric
strong coupling point, either at quarter-filling or at
half-filling~\cite{Galpin05,Galpin06}.  Because of it, it is unclear
what is the relevant symmetry of the system once interaction asymmetry
is present; the unitarity limit of the conductance enhancement at
quarter-filling cannot distinguish between them.
In the standard spin-degenerate $SU(2)$ Kondo systems, experiments
have confirmed universal scaling regarding
temperature~\cite{Goldhaber-Gordon98b} as well as the bias voltage at
odd valleys~\cite{Grobis08,Rincon09}.  Since the absence of the exact
$SU(4)$ symmetry exhibits the behavior at finite temperature and/or at
finite bias voltage, we will try to identify the relevant
universality, $SU(4)$ or $SU(2)\times SU(2)$, by examining the
universal scaling of the linear/nonlinear conductance.
We will show numerically no crossover observed in the conductance
for a experimentally relevant parameters $U'/U=0.4\sim 1.0$ and
$U=10\sim 20 \Gamma$; the Kondo-like enhancement is controlled by
the universal behaviors of the $SU(4)$ Anderson model only with the
characteristic temperature renormalized.

\paragraph*{Outline of the paper}

The paper is organized as follows.  In
Sec.~\ref{sec:model-and-formulation}, we introduce the model of
quantum transport through a nanostructure with two-fold degenerate
orbits, and main apparatus of our approach, finite interaction
slave-boson mean field theory; a brief theory of the linear/nonlinear
conductance is summarized.  In
Sec.~\ref{sec:linear-and-nonlinear-conductance}, we present our
theoretical results both for the linear and nonlinear conductance.  We
see the importance of the interaction asymmetry on the conductance
profile.  A direct comparison with numerical RG data is also made.
After introducing the characteristic temperature $T^{*}$ in
Sec.~\ref{sec:characteristic-temperature}, we discuss our theoretical
results in light of $T^{*}$, and universal scalings of linear and
nonlinear conductance is examined in Sec.~\ref{sec:universal-scaling}.
After discussing an issue of $T^{*}$ renormalization versus a
crossover in Sec.~\ref{sec:Tstar-vs-crossover}, we compare and discuss our
results with typical experiments where the interaction asymmetry is
expected to play an important role in
Sec.~\ref{sec:experimental-manifestation}.  Finally we conclude in
Sec.~\ref{sec:conclusion}.  Part of the results on the linear
conductance has been reported earlier~\cite{Oguchi09}.

\section{Model and Formulation}
\label{sec:model-and-formulation}

\subsection{Quantum dots with doubly degenerate orbitals}

Relatively large energy scale of the Coulomb interaction allows us to
focus on the topmost electron shell.  Our basic assumption is that
within the universality of the topmost shell, the Anderson impurity
model can describe appropriately quantum transport phenomena for any
value of $N_{d}$.
We model the topmost shell of the dot as degenerate in orbits
$i=1,2$. Such degeneracy may come from the coupled-dot or a
unique electronic structure of carbon nanotubes.  

On the dot, an electron
interacts with an electron in the same orbit $i=1, 2$ by $U$ or in the
different orbit by $U'$. The total Hamiltonian is given by 
\begin{equation}
  H = H_{D} +  H_{L} + H_{T},
\label{eq:Anderson-model}
\end{equation}
where the dot $H_{D}$, the noninteracting leads
$H_{L}$, and the coupling between leads ($\alpha=L,R$) and the dot $H_{T}$ are
defined by
\begin{eqnarray}
  && H_{D} = \sum_{i\sigma}  \varepsilon_{d}\, \hat{n}_{i\sigma} + \sum_{i}
    U \hat{n}_{i\uparrow} \hat{n}_{i\downarrow} + 
  U'\hat{n}_{1}\hat{n}_{2}, \\
  && H_{L} = \sum_{\alpha}\sum_{ki\sigma} \left( \varepsilon_{k} -
    \mu_{\alpha} \right) c^{\dagger}_{\alpha ki\sigma}c_{\alpha ki\sigma},\\
  && H_{T} = \sum_{\alpha}\sum_{ki\sigma} \left(t_{k}
    c^{\dagger}_{\alpha k i \sigma}d_{i \sigma}+\text{h.c.} \right).
\end{eqnarray}
Here the number operator of the dot is defined by $\hat{n}_{i} =
\sum_{\sigma} \hat{n}_{i\sigma} = \sum_{\sigma} d^{\dagger}_{i\sigma}
d_{i\sigma}$ and the gate voltage $\varepsilon_{d}$ control
the average electron number on the dot $N_{d}
=\langle \sum_{i}\hat{n}_{i} \rangle$ continuously from
$0$ to $4$.
In the calculation below, we assume constant density of states
$\rho_{\alpha}$ of the lead $\alpha$ within the energy band $(-D,D)$, and use
$\Gamma=\sum_{\alpha}\Gamma_{\alpha} =\sum_{\alpha} \pi \rho_\alpha
|t_k|^2$ as a coupling parameter between the leads and the dot.
To investigate the nonlinear conductance in the presence of finite
source-drain voltage $V$, we choose the chemical
potentials $\mu_{\alpha}$ as $\mu_{L}=-\mu_{R}=eV/2$.

In the case of $U'=U$, the total Hamiltonian $H$ exhibits the full
$SU(4)$ symmetry.  The states $(n_{1},n_{2})=(1,0)$ and $(0,1)$ are
four-fold degenerate at quarter-filling ($N_{d}=1$);
$(n_{1},n_{2})=(2,0)$, $(1,1)$ and $(0,2)$, six-fold degenerate at
half-filling ($N_{d}=2$).  

When one breaks the $SU(4)$ symmetry by decreasing $U'$, the four-fold
degeneracy at quarter-filling is unbroken, but the six-fold degeneracy
at half-filling is broken.  The effect of the interaction asymmetry
appears as shifts of the Coulomb blockade peaks, those at
$\varepsilon_{d}=0$, $-U'$, $-U-U'$, and $-U-2U'$.  However, the effect
of the interaction asymmetry does not stop here; it gives a
substantial effect in the Kondo effect particularly at half-filling.
This simple argument also indicates that the $SU(4)$ symmetry at
half-filling is more vulnerable than that at quarter-filling.  We will
show below that this is indeed the case.

\subsection{Finite-interaction slave-boson mean-field theory}

We analyze the model by an extension of the Kotliar-Ruckenstein
formulation of slave-boson mean field theory
(KR-SBMT)~\cite{Kotliar86}, where a bosonic field is attached to each
type of local excitations.~\cite{footnote1}
The approach has several advantages that
other slave-boson cousins miss. By retaining finite interaction, it
can treat the mixed-valence regime; it applies to nonlinear transport
through a quantum dot~\cite{Dong01}; it reproduces Fermi liquid
behavior at $T=0$ with satisfying the Friedel sum rule; it can
evaluate the linear/nonlinear conductance for the entire range of the
gate voltage systematically.
The KR formulation of SBMT is believed to be an analytical yet reliable
non-perturbative approximation up to the Kondo temperature,
agreeing successfully with numerical RG methods and
experiments~\cite{Dong01,Takahashi06} (see also
Sec.~\ref{sec:linear-and-nonlinear-conductance}). 

When the KR-SBMT is extended to the dot with doubly degenerate orbitals, 16
bose fields are needed associated to each state of the dot: $e$ for
the empty, $p_{i\sigma}$ for one electron with orbit $i$ and spin
$\sigma$, $x_{i}$ for two electrons on the same orbit $i$, $y_{sm}$
for two electrons at different orbits with total spin $(s,m)$,
$h_{i\sigma}$ for three electrons with a hole on $i\sigma$, and $b$
for fully occupied state.  Below we briefly summarize our
approximation scheme (see also ref.~\citen{Dong02}).

As a standard treatment of the Kotliar-Ruckenstein formulation, we
can rewrite the Hamiltonian eq.~\eqref{eq:Anderson-model} in
terms of these newly defined fields by adding the Lagrange multiplier
terms $H_{\Lambda}= \lambda^{(1)} (\mathcal{I} - 1) + \sum_{i\sigma}
\lambda^{(2)}_{i\sigma} \left( d^{\dagger}_{i\sigma} d_{i\sigma} -
  Q_{i\sigma} \right)$ to respect two constraints: the completeness
condition,
\begin{align}
& \mathcal{I} \equiv e^{\dagger}e+\sum_{i\sigma}p^{\dagger} _{i\sigma}
	p_{i\sigma}+\sum_{i}x^{\dagger} _i x_i
\nonumber \\ & \quad
+\sum_{sm}
	y^{\dagger}_{sm}y_{sm} +\sum_{i\sigma}h^{\dagger}_{i \bar{\sigma}} h_{i
  \bar{\sigma}}+b^{\dagger}b = 1, 
\end{align}
and the number correspondence between fermions and newly
defined bosons,
\begin{align}
& \mathcal{Q}_{i\sigma} \equiv p^{\dagger} _{i\sigma}p_{i\sigma}
	+x^{\dagger} _ix_i+y^{\dagger} _{1,2\sigma}y_{1,2\sigma}
	+\frac{1}{2}(y^{\dagger}_{00}y_{00}+y^{\dagger}_{10}y_{10})
\nonumber \\ & \quad
  + h^{\dagger} _{i \bar{\sigma}} h_{i \bar{\sigma}}+\sum_{\sigma'}
	h^{\dagger} _{\bar{i} \sigma'}h_{\bar{i} \sigma'}
	+b^{\dagger}b = d^{\dagger}_{i\sigma} d_{i\sigma}.
\end{align}
Here and hereafter, we adopt the notation $\bar{1}=2$, $\bar{2}=1$,
$\bar{\uparrow}=\downarrow$, $\bar{\downarrow} = \uparrow$.  As a
result, we can rewrite eq.~\eqref{eq:Anderson-model} exactly as
\begin{align}
& H = H_{L}	+
\sum_{i\sigma}\varepsilon_{d}\, d^{\dagger}_{i\sigma} d_{i\sigma}
	+ U\sum_{i} x^{\dagger}_{i}x_{i} +U'\sum_{sm} y^{\dagger}_{sm}y_{sm} 
\notag \\ & \quad
 + (U+2U') \sum_{i\sigma}h^{\dagger}_{i\sigma}h_{i\sigma} +
 (2U+4U')b^{\dagger}b  
\notag\\ & \quad
	+\sum_{k i \sigma} \left( t_{k} c^{\dagger}_{k i \sigma}
	d_{i \sigma}z_{i\sigma}+\text{h.c.} \right) + H_{\Lambda},
\label{eq:SB-exact-H}
\end{align}
where $z_{i\sigma} = (1-Q_{i\sigma}) ^{-\frac{1}{2}} [ e^{\dagger}
p_{i\sigma} + p^{\dagger}_{i \bar{\sigma}}x_i+p^{\dagger} _{\bar{i}
  \sigma}y_{1,2\sigma} + p^{\dagger}_{\bar{i} \bar{\sigma}}
(y_{00}+y_{10})/{2} + x^{\dagger}_{\bar{i}} h_{i \bar{\sigma}} +
(y^{\dagger}_{00} + y^{\dagger}_{10}) h_{\bar{i} \bar{\sigma}}/2 +
y^{\dagger}_{1,2\bar{\sigma}} h_{\bar{i} \sigma} + h^{\dagger}_{i\sigma}
b ] Q_{i\sigma}^{-\frac{1}{2}}$ denotes a renormalization of
one-particle annihilation.  

  By applying the mean field approximation, we replace all the boson
  fields by their expectation values.  We determine those expectation
  values self-consistently by solving equations of motion and
  constraints at each gate voltage $\varepsilon_{d}$, the bias
  condition $V$ and the temperature $T$.  By adopting a symmetric
  solution regarding the orbit $i$ and the spin $\sigma$, we should
  solve the following self-consistent equations (omitting
  spin/orbit indices),
\begin{eqnarray}
	&&4\frac{\partial\ln z}{\partial e}M+\lambda^{(1)}\, e=0,\\
	&&\frac{\partial\ln z}{\partial p}M+(\lambda^{(1)}-\lambda^{(2)})p=0,\\
	&&2\frac{\partial\ln z}{\partial x}M+(U+\lambda^{(1)}-2\lambda^{(2)})x=0,\\
	&&\frac{\partial\ln z}{\partial y}M+(U'+\lambda^{(1)}-2\lambda^{(2)})y=0,\\
	&&\frac{\partial\ln z}{\partial h}M+(U+2U'+\lambda^{(1)}-3\lambda^{(2)})h=0,\\
	&&4\frac{\partial\ln z}{\partial b}M+(2U+4U'+\lambda^{(1)}-4\lambda^{(2)})b=0,
\end{eqnarray}
as well as two constraints
\begin{align}
& e^{2} + 4p^{2} + 2x^{2} + 4y^{2} + 4h^{2} + b^{2} = 1,\\
& N_{d}/4=p^2+x^2+2y^2+3h^2+b^2.
\end{align}
Here we have introduced (non-equilibrium) Green functions $
G^{<}_{d}(t,t') \equiv i \langle
d^{\dagger}_{i\sigma}(t')d_{i\sigma}(t)\rangle$, its Fourier
transformation $G^{<}_{d}(\omega)$, and $M=\int \frac{d\omega}{2\pi
  i}(\omega-\tilde{\varepsilon}_d) G^{<}_d(\omega)$.

\subsection{Current formula and nonlinear conductance}

On determining these auxiliary parameters self-consistently at each
temperature and each gate voltage, the system reduces to the
renormalized resonant level model 
\begin{equation}
  H_{\text{eff}} = H_{L} + \sum_{i} \tilde{\varepsilon}_d\,
  \hat{n}_{i} + \sum_{\alpha k i \sigma} \left( \tilde{t}_{k}\,
    c^{\dagger}_{\alpha k i\sigma}d_{i \sigma}+\text{h.c.} \right),
\label{eq:Heff}
\end{equation}
with the effective dot level
$\tilde{\varepsilon}_{d} = \varepsilon_{d} + \lambda^{(2)}$ and the
effective hopping $\tilde{t}_{k} =z t_{k}$, which corresponds to the effective
peak width $\tilde{\Gamma}= \tilde{\Gamma}_{L} + \tilde{\Gamma}_{R} =
\sum_{\alpha}\pi \rho_{\alpha} |\tilde{t}_{k}|^2$.
The effective Hamiltonian conforms to Fermi liquid description at low
temperature, the strong-coupling fixed point.  Note that interaction
effect is considered only through $\tilde{\varepsilon}_{d}$,
$\tilde{t}_{k}$ and we ignore a remaining (renormalized) interaction
term between quasiparticles.

The form of $H_{\text{eff}}$ enables us to find the linear/nonlinear
conductance by the Meir-Wingreen formula~\cite{Wingreen94}:
\begin{align}
& I = \frac{16e\tilde{\Gamma}_{L}\tilde{\Gamma}_{R}}{h}\int d\omega\,
\frac{\left[ f_{L}(\omega) - f_{R}(\omega) \right]}
{(\omega-\tilde{\varepsilon}_{d})^{2} + \tilde{\Gamma}^{2}}
\\ & \quad
= \frac{4G_{0}\tilde{\Gamma}}{e} \Im\left[
  \psi\left( \tfrac{1}{2} + i \zeta_{R} \right)
-   \psi\left( \tfrac{1}{2} + i \zeta_{L} \right) \right],
\end{align}
where $f_{\alpha}(\omega)=1/[e^{(\omega-\mu_{\alpha})/T}+1]$ is the
Fermi distribution function of the lead, $G_{0} = (4\tilde{\Gamma}_{L}
\tilde{\Gamma}_{R}/\tilde{\Gamma}^{2}) (e^{2}/h)$, $\zeta_{\alpha} =
(\tilde{\varepsilon}_{d} -\mu_{\alpha} - i\tilde{\Gamma})/(2\pi T)$,
and $\psi$ is the digamma function.  By this, we can readily evaluate
the nonlinear conductance at finite bias voltage $V$ by
\begin{equation}
  G(T,V) = \frac{dI}{dV}.
\end{equation}
We should bear in mind, however, that the renormalized parameters
$\tilde{\varepsilon}_{d}$ and $\tilde{\Gamma}$ still depend on $V$ as
well as $T$ and interaction strengths.

At $T=0$, the electron number on the dot is equal to $N_{d} = 2-
\frac{4}{\pi} \arctan \big( \tilde{\varepsilon}_{d}/\tilde{\Gamma}
\big)$.  This immediately shows that the linear conductance at zero
temperature $G(0,0)$ satisfies 
\begin{equation}
	G(0,0) =  4G_{0}\sin^{2}(\pi N_d/4). 
	\label{eq:Friedel-sum-rule}
\end{equation}
This is the Friedel sum rule of the $SU(4)$ Anderson model. The formula
immediately indicates that the zero-bias conductance in the unitarity
limit approaches $G(0,0)\to 2G_{0}$ at $N_{d}=1,3$, but $G(0,0)\to
4G_{0}$ at $N_{d}=2$.
The observation of the above dependence of $G(0,0)$ on continuous
$N_{d}$ is a hallmark of the $SU(4)$ Anderson model; a recent
experiment~\cite{Makarovski07} and calculations by numerical
RG~\cite{Anders08} confirmed it beautifully.

\section{Linear and nonlinear conductance}
\label{sec:linear-and-nonlinear-conductance}

In this section, we present our main numerical results of
linear/nonlinear conductance as a function of the temperature and/or
bias voltage.
The Friedel sum rule indicates the
zero-bias conductance at $T=0$ does not depend on interaction
asymmetry, taking a \emph{universal} form,
eq.~\eqref{eq:Friedel-sum-rule}.  
Accordingly, the absence of the exact $SU(4)$ symmetry appears at finite
temperature and/or bias-voltage; we will find that the effect is
large enough to modify the conductance profile substantially,
providing a characteristic `dip structure' around half-filling.
For all the calculations, we fix a parameter $U/\Gamma=20$ as a
typical value of strong correlation in experiments, and we take $U$ as
a unit energy scale, if needed.

\subsection{$G(T,0)$ of symmetric interaction $U'=U$; comparison
  with NRG results}
\label{sec:zero-bias}

As a first step, we compare the zero-bias conductance $G(T,0)$
obtained by our KR-SBMT approach with the NRG
calculations available for the $SU(4)$ Anderson model
with symmetric interaction $U'=U$~\cite{Anders08}.  We find that our results reproduce
NRG data quite well.

\begin{figure}
	\begin{center}
    \includegraphics[width=0.9\linewidth]{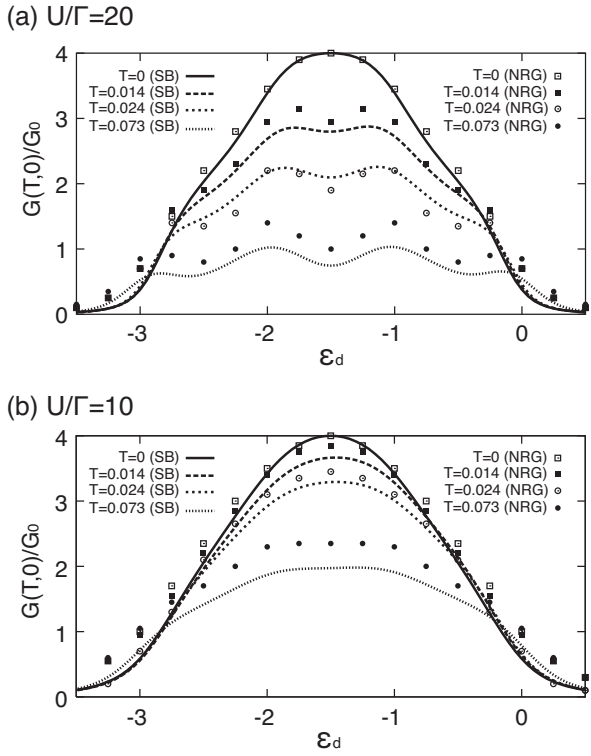}
	\end{center}
	\caption{Comparison between the KR-SBMT and numerical RG
    results~\cite{Anders08}.  Temperature evolution of linear
    conductance $G(T,0)$ is plotted as a function of the gate voltage
    for the fully symmetric $SU(4)$ Anderson model.  (a) SBMT results
    (lines) of $U/\Gamma=20$ are compared with NRG data (symbols) of
    $U/\Gamma=10$; (b) SBMT results (lines) of $U/\Gamma=10$, with NRG
    data (symbols) of $U/\Gamma=5$ (see the text for details).}
	\label{fig:G-comp-NRG}
\end{figure}

Figure~\ref{fig:G-comp-NRG} (a) demonstrates a distinctive feature of the
conductance profile in large $U/\Gamma$ region, the Kondo enhancement
both at $N_{d}\approx 2$ and at $N_{d}\approx 1$; with increasing
temperature, the four-fold Coulomb blockade peaks appear.  
Regarding the mixed-valence regime with a smaller value of $U/\Gamma$,
four-fold peaks merge to form one big peak showing the Kondo enhancement
(see Fig.~\ref{fig:G-comp-NRG} (b)).  As was claimed
already~\cite{Anders08}, these temperature evolutions by $SU(4)$
symmetric Anderson model agree very well with what is observed either
at $V_{g} \sim 3.9\textrm{V}$ or at $V_{g} \sim 5.3\textrm{V}$ in an
experiment~\cite{Makarovski07}.

Our KR-SBMT results agree even quantitatively well with NRG results,
but with one catch:  In comparing with NRG data, we need to resort to a 
heuristic prescription choosing the twice as large value of $U/\Gamma$ for the
 SBMT results.~\cite{footnote2}  
 Indeed, the same problem has been prevailing in the single impurity
 Anderson model; we can attain quantitative agreement between
 effective theories (the KR-SBMT~\cite{Takahashi06} or the functional
 RG~\cite{Karrasch06}) and the NRG method only if we choose the twice
 as large value of $U/\Gamma$ for the former.
 Figure~\ref{fig:G-comp-NRG} shows comparison by using this heuristic
 prescription, and they agree very well in low temperature, roughly up to the
 Kondo temperature scale. 

\subsection{Effect of interaction asymmetry on $G(T,0)$}
\begin{figure}
	\begin{center}
    \includegraphics[width=0.7\linewidth]{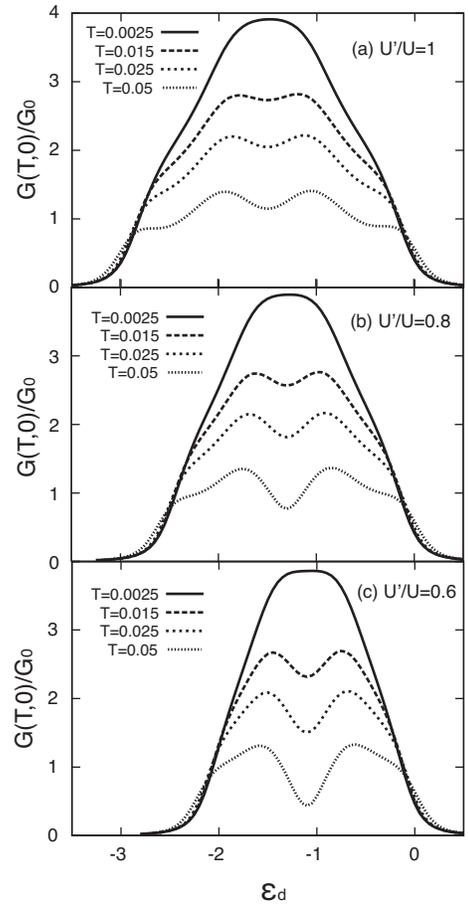}
	\end{center}
	\caption{The temperature evolutions of the conductance $G(T,0)$ as a
    function of the gate voltage with varying asymmetry $U'/U=1$,
    $0.8$, and $0.6$.}
	\label{fig:GT0-Ed}
\end{figure}

Figure~\ref{fig:GT0-Ed} demonstrates typical temperature
evolutions of the conductance profile as a function of the gate
voltage with varying asymmetry parameters (a)~$U'/U=1$, (b)~$0.8$, and
(c)~$0.6$.  Overall widths of conductance profiles are determined by
$U+2U'$.  We notice, even with a small asymmetry $U'/U=0.8$, a
characteristic dip structure of $G(T,0)$ developing clearly at
half-filling ($N_{d}=2$) at finite temperature, in contrast with a
small deformation around quarter and three-quarter-fillings
($N_{d}=1,3$).
Recall for any finite $0<U'<U$, the system is always renormalized to
the $SU(4)$ symmetric model~\cite{Galpin05,Galpin06} and the Kondo
enhancement always appear at the gate voltage corresponding to
$N_{d}=1,2,3$; $G(0,0)$ is determined by the Friedel sum rule
eq.~\eqref{eq:Friedel-sum-rule} irrespective of the interaction asymmetry.
We stress the interaction asymmetry manifests itself only at \emph{finite}
temperature and invisible at $T=0$. It appears as a substantial `dip
structure' of $G(T,0)$ around half-filling.  As we have claimed
recently~\cite{Oguchi09}, such feature caused by interaction
asymmetry reproduces quite well experimental observation in carbon
nanotube dots~\cite{Jarillo-Herrero05,Makarovski07} (see 
Sec.~\ref{sec:experimental-manifestation} for more details).

\begin{figure}
	\begin{center}
    \includegraphics[width=0.9\linewidth]{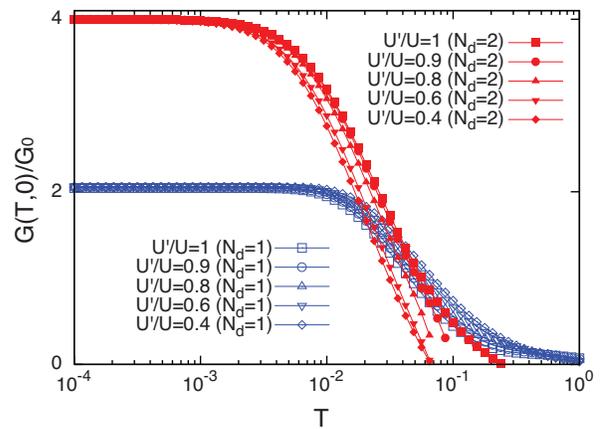}
	\end{center}
	\caption{(color online) Temperature dependence of the conductance $G(T,0)$ as a
    function of the interaction asymmetry $U'/U=1$, $0.9$, $0.8$, $0.6$, and
    $0.4$ at $N_d=1$ (quarter-filling) and at $N_d=2$ (half-filling).}
	\label{fig:GT0-T}
\end{figure}
To clarify how the asymmetric interaction affect the conductance
profile particularly at half-filling, we make a direct comparison
between the behaviors at $N_{d}=1$ and $N_{d}=2$ in
Fig.~\ref{fig:GT0-T} by changing $U'/U=1$, $0.9$, $0.8$, $0.6$, and
$0.4$.  By decreasing $U'$ away from the symmetric point $U'=U$, the
characteristic energy controlling thermal suppression seems reduced at
$N_{d}=2$, but enhanced at $N_{d}=1$.  We will elucidate this
tendency by estimating $T^{*}$ numerically in
Sec.~\ref{sec:characteristic-temperature}.

\subsection{Effect of interaction asymmetry on $G(0,V)$}
\label{sec:finite-bias}
We now turn our attention to the non-equilibrium transport, {\it
  i.e.}, nonlinear conductance.  To understand the interplay
between finite bias voltage effect and the interaction asymmetry, we
here examine the zero-temperature differential conductance $G(0,V)$.

\begin{figure}
	\begin{center}
    \includegraphics[width=0.7\linewidth]{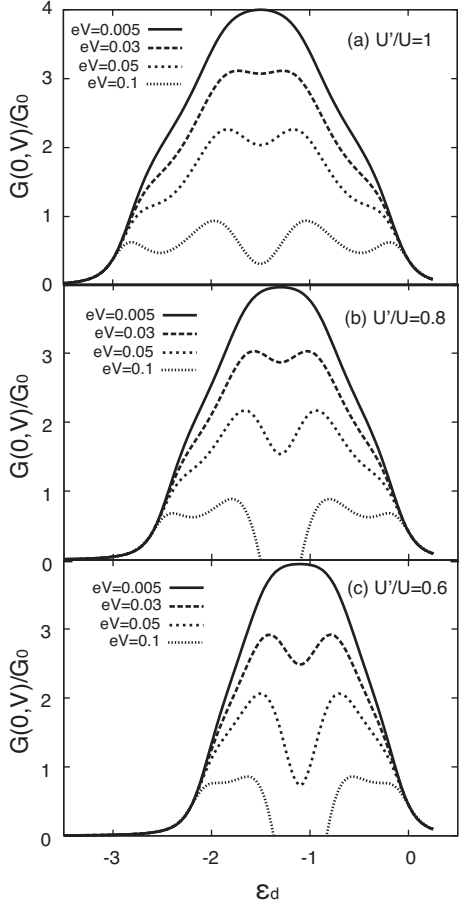}
	\end{center}
	\caption{Bias voltage evolution of the zero-temperature differential
    conductance $G(0,V)$ as a function of the gate voltage with varying
    the interaction asymmetry (a) $U'/U=1$, (b) $U'/U=0.8$, and (c) $U'/U=0.6$.}
	\label{fig:G0V-Ed}
\end{figure}
Figure~\ref{fig:G0V-Ed} shows typical bias voltage evolutions of
$G(0,V)$ as a function of the gate voltage with asymmetry parameters
(a)~$U'/U=1$, (b)~$0.8$, and (c)~$0.6$.  We immediately notice that
the profile of $G(0,V)$ looks quite similar to that of $G(T,0)$.  By
applying larger bias voltage ($eV/U=0.03$, $0.05$, $0.1$), the Kondo
enhancement both at $N_{d}=1,3$ and $N_{d}=2$ is suppressed, but the
latter suppression is more distinctive.  Like the thermal evolutions,
the interaction asymmetry manifests itself only at a fixed
\emph{finite} bias voltage, as a `dip structure' of $G(0,V)$ around
half-filling.

\begin{figure}
	\begin{center}
    \includegraphics[width=0.9\linewidth]{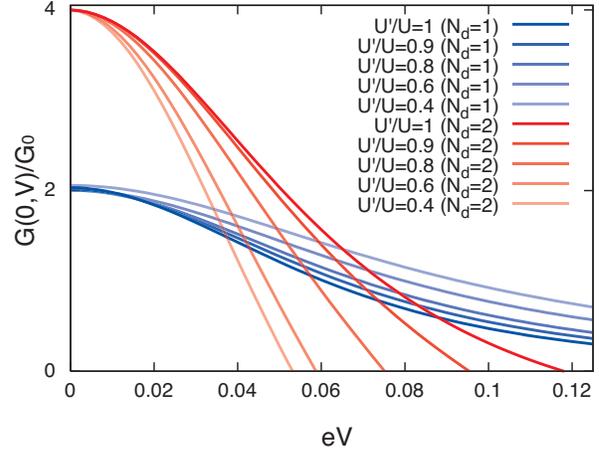}
	\end{center}
	\caption{(color online) The differential conductance $G(0,V)$ as a
    function of the bias voltage with varying the interaction asymmetry
    $U'/U=1, 0.9, 0.8, 0.6, 0.4$  at $N_d=2$ (approaching $G(0,V\to 0)\to
    4G_{0}$), and at $N_d=1$ (approaching $G(0,V\to 0)\to 2G_{0}$).} 
	\label{fig:G0V-V}
\end{figure}
Figure~\ref{fig:G0V-V} shows the zero bias peak $G(0,V)$
either at $N_{d}=1$ or at $N_{d}=2$ with varying the asymmetry
parameter $U'/U=1.0$, $0.9$, $0.8$, $0.6$, and $0.4$.  
Since we assume the symmetric coupling $\Gamma_{L}=\Gamma_{R}$ to the
leads, the $G(0,V)$ becomes symmetric regarding $V$.  From its
behavior, we can read off how the characteristic energy scale depends
on the asymmetric parameter $U'/U$.  The dependence is the same with what is
observed in the temperature evolutions of $G(T,0)$; decreasing $U'$
leads to the reduction of the characteristic energy at $N_{d}=2$, but
the enhancement at $N_{d}=1$.

Our numerical calculations show that the conductance at $N_{d}=2$

\section{Characteristic temperature $T^{*}$}
\label{sec:characteristic-temperature}

The Kondo temperature is an energy scale characterizing the Kondo
effect, {\it i.e.}, the Anderson model eq.~\eqref{eq:Anderson-model}
at the center of an odd Coulomb blockade valley.  It is a crossover
temperature, not the transition temperature, so we can fix it only
modulo numerical factor with ambiguity. 
Instead, we seek to define a characteristic energy scale at each gate
voltage covering the entire range of the topmost shell.

Recalling the present effective theory expresses the dot Green function as
\begin{equation}
  G^{R}_{d}(\omega=0) =
  \frac{1}{-\tilde{\varepsilon}_{d}+i\tilde{\Gamma}} = -
\frac{\exp(i\pi N_{d}/4)}{\sqrt{\tilde{\varepsilon}_{d}^{2} +
    \tilde{\Gamma}^{2}}}, 
\end{equation}
we naturally define a characteristic temperature $T^{*}$ of the model
\begin{equation}
	T^{*} = \sqrt{\tilde{\varepsilon}_{d}^{2} + \tilde{\Gamma}^{2}}\,
	\Big|_{T=0,V=0}
	\label{eq:T-star}
\end{equation}
at \emph{each} gate voltage or $N_{d}$.
We stress it is imperative to include the shift
$\tilde{\varepsilon}_{d}$ in defining a consistent energy scale at
each $N_{d}$. 

When $\tilde{\varepsilon}_{d}$ vanishes, we see
$T^{*}=\tilde{\Gamma}$, the Kondo peak width.  This is a situation of
the $SU(2)$ Kondo effect.  However, as derived from the Friedel sum
rule of $SU(4)$ model, eq.~\eqref{eq:Friedel-sum-rule}, the $SU(4)$
Kondo peak needs to shift at the $N_{d} =1,3$~\cite{Choi05,Lim06} by
$|\tilde{\varepsilon}_{d}| = \tilde{\Gamma} =T^{*}/\sqrt{2}$.  In contrast,
the $SU(4)$ Kondo peak at $N_{d}=2$ does not shift because of the
electron-hole symmetry.  

\begin{figure}
	\begin{center}
    \includegraphics[width=0.9\linewidth]{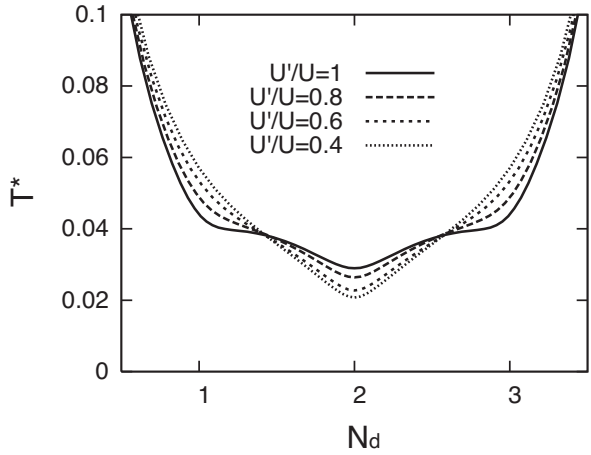}
	\end{center}
	\caption{Dependence of the characteristic temperature $T^*$ 
    on the average electron number of the dot with varying interaction
    asymmetry $U'/U$.}
	\label{fig:Tstar-Nd}
\end{figure}
\begin{figure}
	\begin{center}
    \includegraphics[width=0.9\linewidth]{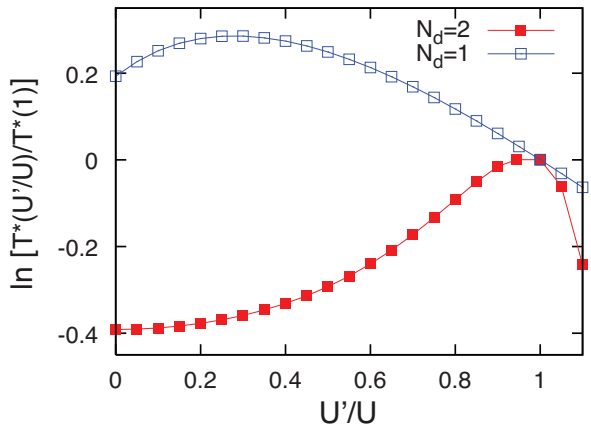}
	\end{center}
	\caption{(color online) Dependence of the characteristic temperature
    $T^*$ on the interaction asymmetry $U'/U$ for $N_d=2$ (filled
    square) and for $N_d=1$ (empty square).}
	\label{fig:Tstar-Uprime}
\end{figure}
The dependence of the characteristic energy $T^{*}$ obtained from our
KR-SBMT calculations is shown in Fig.~\ref{fig:Tstar-Nd} for the
entire range of the topmost shell ($N_{d}=0 \sim 4$).  Though the
state degeneracy at $U'=U$ is six-fold at $N_{d}=2$ and four-fold at
$N_{d}=1$ respectively, $T^{*}$ of the former is lower than that of
the latter. 

The presence of the interaction asymmetry further amplifies this
difference of $T^{*}$ as in Fig.~\ref{fig:Tstar-Uprime}; it reduces
$T^{*}$ at $N_{d} =2$ but enhances it at $N_{d}=1$; $T^{*}$ is almost
unchanged at the filling corresponding to Coulomb blockade peaks
($N_{d} \approx 1.3$ and $2.7$ in Fig.~\ref{fig:Tstar-Nd}).  Because
of this, the Kondo enhancement around $N_{d}\approx 2$ is destroyed
more rapidly than at $N_{d}\approx 1$ either by finite temperature or
by finite bias voltage.  The above dependence of $T^{*}$ gives a
consistent explanation on the conductance profiles
Figs.~\ref{fig:GT0-Ed}--\ref{fig:G0V-V} in the previous section.

How much asymmetry deforms the conductance profile of the $SU(4)$
symmetry?  The characteristic temperature $T^{*}$ at half-filling (of
the symmetric model) controls it.  When $|U-U'|$ exceeds
$T^{*}(N_{d}=2, U'=U)$, it reduces $T^{*}$ rapidly, producing the
conductance profile with a distinctive dip at half-filling. One can
observe the effect only at finite temperature or by finite bias
voltage.  The above behavior of $T^{*}$ at $N_{d}=1$, $2$ agrees
qualitatively with other numerical estimates of the Kondo
temperature~\cite{Galpin05,Galpin06,Mitchell06,Busser07}.  However, we
need to distinguish between the reduction of the characteristic energy
and the crossover between the universality classes: the former does
not necessarily mean the latter.  We will focus on the issue in next section.

\section{Universal scaling of linear and nonlinear conductance}
\label{sec:universal-scaling}

Having understood qualitative behavior of the conductance profiles by
temperature or bias voltage, we now focus on examining an important
issue: what is the universality controlling quantum transport in the
presence of the interaction asymmetry?  The nature at half-filling
have been argued actively so far, proposing the $SU(2) \times SU(2)$
Kondo effect~\cite{Galpin05,Galpin06,Mitchell06}, or the $SU(4)$ Kondo
effect~\cite{Busser07}.  We should bear in mind that the absence of
the exact $SU(4)$ symmetry of the model does not necessarily
legitimate the former universality, because the model flows toward the
$SU(4)$ symmetric strong coupling point for $0<U'<U$; neither the
reduction of $T^{*}$ in Fig.~\ref{fig:Tstar-Uprime}.

When the nonlinear conductance $G(T,V)$ is dominated by a unique
universality, we expect to express $G(T,V)$ by some universal function
$F_{n}(t=T/T^{*},v=eV/T^{*})$ at each filling $n=N_{d}$:
\begin{equation}
G(T,V)/G_{0} = F_{n}(t,v).
\end{equation}
The Friedel sum rule eq.~\eqref{eq:Friedel-sum-rule} constrains $F_{n}(0,0)
= 4 \sin^{2}(\pi n /4)$.  
Otherwise, with a crossover present between the universality classes,
we should observe a continuous change of the universal function
$F_{n}(t,v)$.  We test numerically the universality by making
universal scalings of the temperature evolutions and the bias voltage
evolutions of the conductance.

In the following, we will find the temperature or bias-voltage
evolutions of the conductance $G(T,V)$ indeed exhibits itself of the
$SU(4)$ behavior for different values of $U'/U<1$.  Hence we will
claim the phenomena is dominated solely by the $SU(4)$ universality,
\emph{not} a crossover between two different universality classes.

\subsection{Universal scaling of $G(T,0)$}

\begin{figure}
	\begin{center}
    \includegraphics[width=0.9\linewidth]{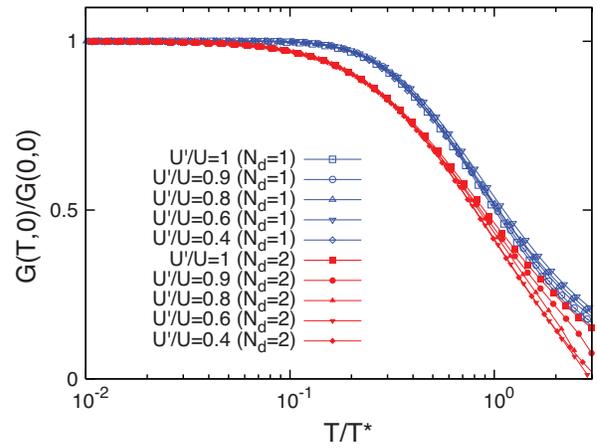}
	\end{center}
	\caption{(color online) Universal scaling of the temperature scaled by
    $T^{*}$. Data are the same with Fig.~\ref{fig:GT0-T}.}
	\label{fig:GT0-scaledT}
\end{figure}
We first examine a universal form of the temperature evolutions of the
linear conductance
\begin{equation}
  G(T,0)/G_{0} = F_{n}(t, 0); \quad t=T/T^{*}.
\end{equation}
In Fig.~\ref{fig:GT0-scaledT}, universal temperature dependence of
$G(T,0)$ either at $N_{d}=1$ or $N_{d}=2$ are confirmed numerically,
by rescaling Fig.~\ref{fig:GT0-T} as a function of $t=T/T^{*}$ with
varying $U'/U=0.4 \sim 1.0$.  Here we defined $T^{*}$ at each $N_{d}$
and $U'/U$.  The results are striking.  For all values of
$U'/U$ both at $N_{d}=1,2$, universal curves collapse well up to
$t\lesssim 1$ (an upper scale restricting the validity of the present
analysis).  It implies that the universal temperature dependence
either at half-filling or at quarter-filling reduces to the $SU(4)$
symmetric case.  Additionally Fig.~\ref{fig:GT0-scaledT} shows that
such universal dependence differs slightly but significantly between
$N_{d}=1,2$.  The universal curves obtained here look very similar to
those of the symmetric model $U'=U$ obtained in NRG~\cite{Anders08}.

\subsection{Universal scaling of $G(0,V)$}

\begin{figure}
	\begin{center}
    \includegraphics[width=0.9\linewidth]{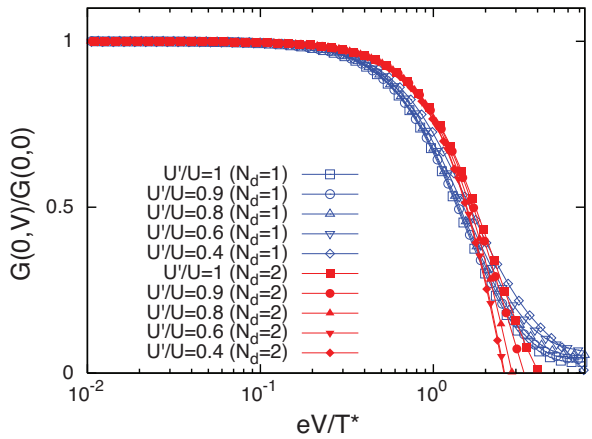}
	\end{center}
	\caption{(color online) Universal scaling of the bias voltage scaled by
    $T^{*}$. Data are the same with Fig.~\ref{fig:G0V-V}.}
	\label{fig:G0V-scaledV}
\end{figure}

We now turn our attention to universal scaling of non-equilibrium
transport regime, of the nonlinear conductance 
\begin{equation}
  G(0,V)/G_{0} = F_{n}(0,v); \quad v=eV/T^{*}. 
\end{equation}

In Fig.~\ref{fig:G0V-scaledV}, we show the bias-voltage evolution
$G(0,V)$ as a function of the scaled bias voltage $v=eV/T^{*}$ either
at $N_{d}=1$ or at $N_{d}=2$.  As the most important observation, we
can confirm that all of the universal scaling curves for
$U'/U=0.4\sim 0.9$ collapse on the $SU(4)$ symmetric one, similarly
to temperature dependence.  In comparing with thermal universal curves
$F_{n}(t,0)$ in Fig.~\ref{fig:GT0-scaledT}, bias-voltage universal
curves $F_{n}(0,v)$ at $N_{d}=1,2$ exhibit rather similar behavior, 
up to $t \lesssim 1$ ($T\lesssim T^{*}$).

\subsection{Universal scaling of $G(T,V)$}

As a concrete dependence, thermal universal curve $F_{n}(t,0)$ and
bias-voltage one $F_{n}(0,v)$ behave differently for $n=N_{d}=1,2$.  
This means that we need three independent variables $t$, $v$, and $n$
fully to describe a universal curve.  However, a simpler description
is possible to characterize the suppression by finite temperature or
bias-voltage.  The idea essentially comes from seeking what is a
cut-off energy of the RG flow.  

\begin{figure}
	\begin{center}
    \includegraphics[width=0.65\linewidth]{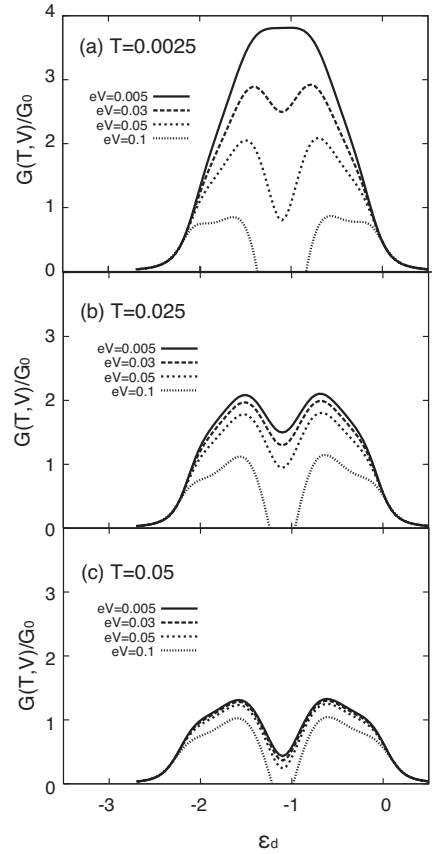}
	\end{center}
	\caption{Conductance profiles $G(T,V)$ by varying different temperature and
    bias voltage.  (a) $T = 0.0025U$ (b) $T=0.025U$ (c) $T=0.05U$. For
    each temperature, results of bias voltage $eV = 0.005$, $0.03$, $0.05$, and
    $0.1U$ are shown.  The characteristic energy at half-filling is
    $T^{*}(2)=0.023U$.  }
	\label{fig:GTV-Ed}
\end{figure}
\begin{figure}
	\begin{center}
	    \includegraphics[width=0.9\linewidth]{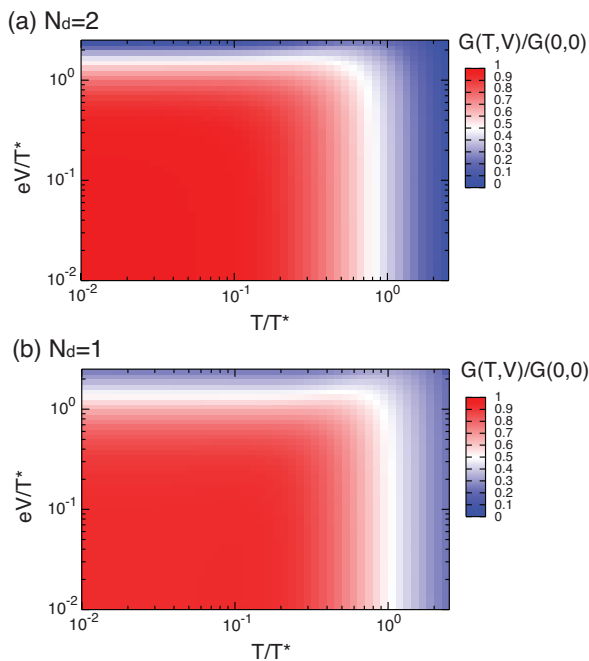}
	\end{center}
	\caption{(color online) Universal function $F_{n}(t,v)$ of the
    $SU(4)$ universality class, obtained by the KR-SBMT. (a)
    $G(T,V)/G(0,0)=F(t,v)/4$ at half-filling ($N_{d}=2$) and (b)
    $G(T,V)/G(0,0)=F(t,v)/2$ at quarter-filling ($N_{d}=1$), }
	\label{fig:Ftv-T-V}
\end{figure}

A typical nonlinear conductance evolution at finite temperature is
shown in Fig.~\ref{fig:GTV-Ed} for $U'/U=0.6$, for (a)~$T=0.0025U$,
(b)~$0.025U$ and (c)~$0.05U$.  In this system, the characteristic
temperature $T^{*}$ at half-filling is estimated as
$T^{*}(N_{d}=2)=0.023U$.  We understand its behaviors by classifying
these behaviors, comparing with this energy scale $T^{*}(2)$.
At (a) $T=0.0025U \ll T^{*}(2)$ ($t\ll 1$), the conductance profile
looks similar to that at $T=0$ in Fig.~\ref{fig:G0V-Ed} (c).  It shows
the temperature is so small that no thermal effect appears.  On the
other hand, at (c) $T =0.05U \gg T^{*}(2)$ ($t\gg 1$), all the
conductance profiles with different bias-voltage $V$ almost collapse
except for $eV=0.1U$ which is larger than the temperature $T=0.05U$,
which is a hallmark of the temperature $T \gg T^{*}(2)$ controlling
the conductance profile for a region $T \gtrsim eV$.  
Consistent behaviors are also confirmed at (b)~$T=0.025U \approx
T^{*}(2)$: the temperature controls the conductance profile for
$T>eV$; the bias voltage, for $T<eV$.  In Fig.~\ref{fig:Ftv-T-V}, we
illustrate universal functions $F_{n}(t,v)$ for $n=1,2$ of the $SU(4)$
universality, obtained by our KR-SBMT approach.  
We conclude that the larger energy scale $\max[t,v]$ between the
temperature and the bias voltage effectively dominates the nonlinear
conductance profile.  The result presented in Fig.~\ref{fig:Ftv-T-V}
is indeed calculated for $U'/U=0.6$, but as we have shown so far, this
universal function $F_{n}(t,v)$ is almost the same up to $T^{*}$ with
that of $U'/U=1$ and also of $U'/U <1$.

Before concluding this section, we give two cautionary comments
regarding the applicability of the present universal dependence of the
conductance: (1)~on conductance profiles observed as a function of the
gate voltage $\varepsilon_{d}$; (2)~on the dependence on the value of
$U/\Gamma$.
The first point is related to our employed approximation (KR-SBMT).
Although we believe the present analysis of the $T^{*}$ scaling may be
useful even for $\max[T,eV] \gtrsim T^{*}$, our present KR-SBMT method
is believed to describe the system adequately only up to $\lesssim
T^{*}$; it does not fully incorporate charge fluctuations, which will
be important for transport under a large bias.  The
characteristic temperature $T^{*}$ is lower at half-filing, higher at
quarter-filling, particularly in the presence of the interaction
asymmetry (Figs.~\ref{fig:Tstar-Nd}, \ref{fig:Tstar-Uprime}).  Hence
the conductance evaluated by the KR-SBMT is validated up to a smaller
value $T$ or $eV$ at half-filling than at quarter-filling.
Such signature is observed in the conductance profile for $eV=0.1U$ in
Figs.~\ref{fig:G0V-Ed} and~\ref{fig:GTV-Ed}, where $\max[T,eV]]\gg
T^{*}(N_{d}=2)$ and the Coulomb blockade physics dominates the
system.  In those cases, the conductance is anticipated to vanish,
with some nonlinear bias effect that the KR-SBMT is likely to miss.
The second point is related to why our analysis observes the $SU(4)$
universality class, disregarding the interaction asymmetry, without a
crossover.  This important issue will be examined further in next
section.

\section{$T^{*}$ renormalization or a crossover: Effect of finite $U$} 
\label{sec:Tstar-vs-crossover}

As we have shown, our results of conductance at finite temperature
and/or bias voltage behave consistently with the $SU(4)$ universality
class even in the presence of the interaction asymmetry $U'/U <1$, 
exhibiting no crossover toward a different universality class.
The results may seem odd at first glance, contradicting with the
suggested crossover between $SU(4)$ and $SU(2)\times SU(2)$
universality classes~\cite{Galpin05}.  However, such a crossover
behavior takes place only for a larger value of $U/\Gamma$.  In a
usual range of experimentally feasible parameters, we claim the
renormalization of the characteristic temperature $T^{*}$, not a
crossover of the universality classes, plays a more dominant role,
controlling the conductance behavior.

\begin{figure}
	\begin{center}
	    \includegraphics[width=0.9\linewidth]{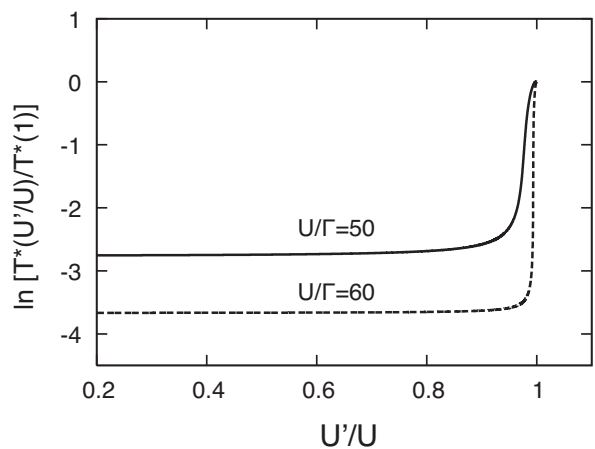}
	\end{center}
	\caption{Dependence of the characteristic temperature
    $T^*$ on the interaction asymmetry $U'/U$ at $N_d=2$ for
    $U/\Gamma=50$ and $60$.}
	\label{fig:Tstar-Uprime-largeU}
\end{figure}

\begin{figure}
	\begin{center}
	    \includegraphics[width=0.9\linewidth]{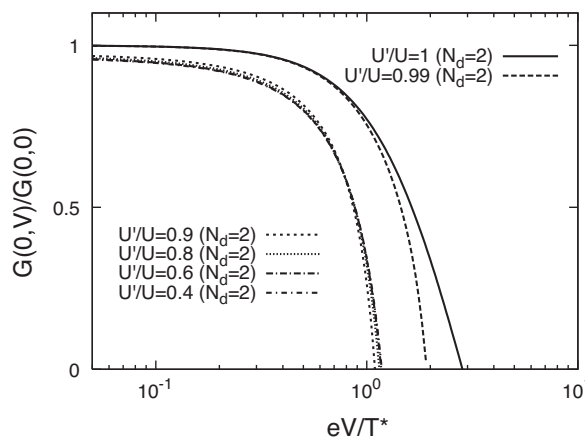}
	\end{center}
	\caption{Universal scaling of the bias voltage scaled by
    $T^{*}$ at $N_{d}=2$ for $U/\Gamma=50$. A crossover between
    different universality classes occurs.}
	\label{fig:G0V-scaledV-largeU}
\end{figure}

To support our view, we also examine the behavior with larger values of
$U$.  Figure~\ref{fig:Tstar-Uprime-largeU} shows the characteristic
temperature $T^{*}$ at $N_{d}=2$ as a function of the interaction
asymmetry $U'/U$ for $U/\Gamma=50, 60$, to be compared with
Fig.~\ref{fig:Tstar-Uprime}.  In this case, the interaction asymmetry
$U'/U$ reduces $T^{*}$ quite rapidly, which signals a crossover of the
universality class from $SU(4)$ to $SU(2)\times
SU(2)$~\cite{Galpin05}.  We confirm that such a crossover indeed
occurs by performing the universal scaling
(Fig.~\ref{fig:G0V-scaledV-largeU}).  For $U/\Gamma=50$, the universal
scaling indicates that the scaled curve for $U'/U=0.9$ is clearly
different from that for $U'/U=1$ or $0.99$; the universality class
crosses over between $SU(4)$ and $SU(2)\times SU(2)$ in this case.

In usual experiments, however, $U$ is roughly in the range of $U = 10
\sim 20\Gamma$ (see also the next section).  For this case, as was
shown in \S\ref{sec:universal-scaling}, the dominant effect at
$N_{d}\approx 2$ is due to the renormalization of $T^{*}$, not a
crossover of the universality classes.

\section{Experimental manifestation}
\label{sec:experimental-manifestation}

We have found theoretically the effect of the interaction asymmetry on the
conductance profile substantial, either at finite temperature or at
finite bias voltage.  Currently, a few experimental realization of the
$SU(4)$ Anderson model eq.~\eqref{eq:Anderson-model} are known:
transport through a carbon nanotube dot, and, through double quantum
dots coupling capacitively with each other~\cite{Wilhelm00,Wilhelm02}.
We discuss our present results in the light of existing experimental
data.

\subsection{Carbon nanotube dot}

Several groups have investigated and observed the Kondo enhancement
in the linear conductance measurement through a carbon nanotube
dot~\cite{Nygard00,Babic04,Jarillo-Herrero05,Makarovski07}.
A carbon nanotube dot has doubly degenerate orbitals in the topmost
shell, and moreover, the conductance according to the $SU(4)$ Friedel
sum rule eq.~\eqref{eq:Friedel-sum-rule} is clearly observed
experimentally at low temperatures~\cite{Makarovski07}. This
observation clearly
justifies the validity of the $SU(4)$ Anderson model.
Notwithstanding, we have not understood fully the temperature evolution
of the conductance, particularly around $N_{d} \approx 2$: While the
experiment by Makarovski \textit{et al.}  has observed visible
low-temperature ``Kondo-like enhancement'' around
$N_{d}=2$\cite{Makarovski07}, such enhancement was absent exhibiting a
characteristic `dip' in others~\cite{Babic04,Jarillo-Herrero05}.
We suggest the interaction asymmetry enables us to understand
systematically finite temperature evolutions in these
experiments~\cite{Oguchi09}.

\begin{figure}
  \begin{center}
    \includegraphics[width=0.9\linewidth]{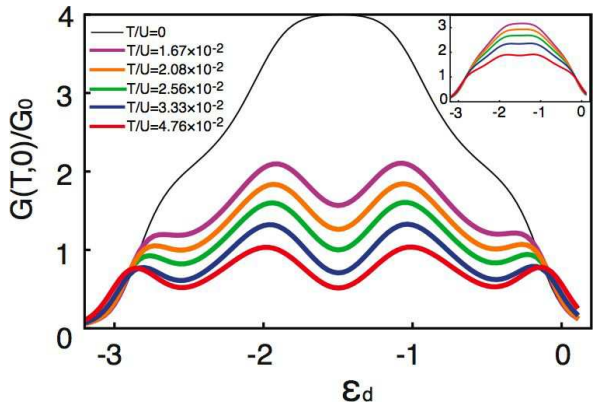}
  \end{center}
  \caption{(color online) Schematic plotting for the comparison with
    carbon nanotube dot experiments: Conductance as a function of the gate voltage for
    $U'/U=0.997$ and $U=30\Gamma$.  $|U-U'| \ll T^{*}(N_{d}=2, U'=U)$
    is realized.  Inset shows larger coupling parameter regime $U =
    15\Gamma$.  }
  \label{fig:small-dU}
\end{figure}
\begin{figure}
  \begin{center}
    \includegraphics[width=0.9\linewidth]{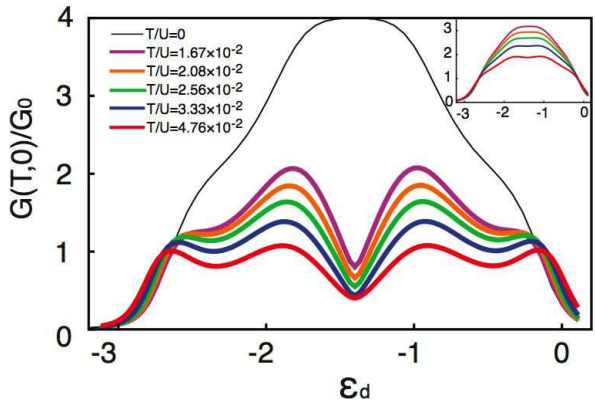}
  \end{center}
  \caption{(color online) Schematic plotting for the comparison with
    carbon nanotube dot experiments: Conductance as a function of the gate voltage for
    $U'/U=0.9$. Other parameters are the same with
    Fig.~\ref{fig:small-dU}.  $|U-U'| \gg T^{*}(N_{d}=2, U'=U)$ is
    realized.}
  \label{fig:large-dU}
\end{figure}

In Figs.~\ref{fig:small-dU} and~\ref{fig:large-dU}, we present
schematic calculations mimicking the experimental situation, changing
$U'/U$ slightly, with $U=30\Gamma$ (with $U=15\Gamma$ for the insets).  
In Fig.~\ref{fig:small-dU} with $U'/U=0.997$, the conductance profile
reproduces all the features of the $SU(4)$ symmetric model.  Notably,
the Kondo enhancement at low temperature occurs either at
$N_{d}\approx 2$ or at $N_{d}\approx 1$ similarly.  As was claimed
already in ref.~\citen{Anders08}, these temperature evolutions by the $SU(4)$
symmetric Anderson model agree very well with what is observed either
at $V_{g} \sim 3.9\textrm{V}$ or at $V_{g} \sim 5.3\textrm{V}$ in
the experiments by Makarovski \textit{et al.}~\cite{Makarovski07}.
To our surprise, however, the conductance profiles with four peaks
($U=30\Gamma$) are modified considerably by a relatively small
asymmetry (less than 10\%), having a characteristic dip structure
around half-filling as in Fig.~\ref{fig:large-dU}.  This is because
the characteristic energy at $N_{d}=2$ is rather small,
$T^{*}(N_{d}=2,U'=U) \approx 0.014U$.  The value $U'/U=0.9$
conforms to almost equally-spaced Coulomb blockade peaks, observed at
$T=8\textrm{K}$~\cite{Jarillo-Herrero05}.  We believe
Fig.~\ref{fig:large-dU} captures essential characteristics
experiments~\cite{Babic04,Jarillo-Herrero05} with a reasonable choice
of physical parameters; for instance, in Fig.~3(a) or Fig.~SI2 of
Jarillo-Herreo \textit{et al}.\cite{Jarillo-Herrero05}, the
conductance at $V_{g} \approx 3\textrm{V}$ is smaller than that of
$V_{g} \approx 2.8\textrm{V}$ or $3.2\textrm{V}$ at each temperature.

Regarding the mixed-valence regime ($U/\Gamma=15$, insets of
Figs.~\ref{fig:small-dU} and~\ref{fig:large-dU}), four peaks merge to form one
big peak showing the Kondo enhancement.  In this regime, we find the interaction
asymmetry $U'/U$  hardly affecting the conductance profile even at
finite temperature. 

\subsection{Double dot system with capacitive coupling}

Several groups have been investigating experimentally on orbital
and/or spin Kondo effects in double quantum dot systems with
capacitive coupling~\cite{Wilhelm00,Wilhelm02,Hubel07,Hubel08}, where
the orbital Kondo effect has been observed at quarter-filling.  The
measured Kondo temperature is found surprisingly high.  Thus we should
possibly attribute the phenomena to the $SU(4)$ Kondo effect, as an
entanglement of spins and orbits.  Therefore, within the universality
of the topmost shell, we will be able to describe the system by
applying the $SU(4)$ Anderson model eq.~\eqref{eq:Anderson-model}.

Although the $SU(4)$ Anderson model suggests that the conductance
should exhibit the Kondo enhancement both at quarter-filling and at
half-filling, to the best of our knowledge, no enhancement has been
observed so far at half-filling.  However, we argue that the absence
of the Kondo-like enhancement at half-filling is due to the combined
effect of finite bias effect and the interaction asymmetry in those
systems.  Consequently, it is expected that the conductance
enhancement should appear also at half-filling by suppressing the bias
voltage at low temperature.

\begin{figure}
	\begin{center}
    \includegraphics[width=0.9\linewidth]{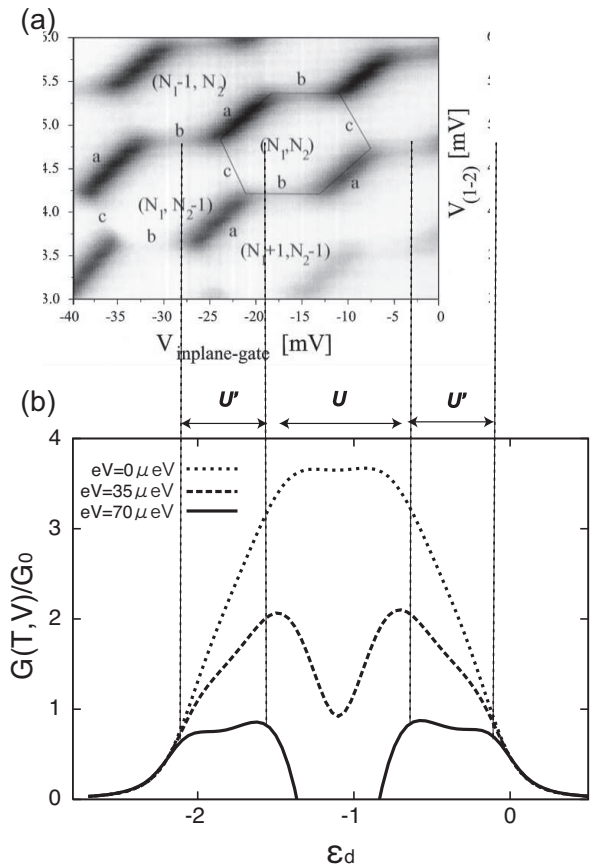}
	\end{center}
	\caption{(a) Experimental $IV$ characteristics excerpted
    from refs.~\citen{Wilhelm00,Wilhelm02} (see the text). (b)~A 
    schematic differential conductance obtained by the KR-SBMT at
    $T=30\,\textrm{mK}$, as
    a function of the gate voltage, varying bias
    voltage $V=0$, $35$, $70\, \mu\textrm{V}$.}
	\label{fig:comp-DQD}
\end{figure}

To make our argument more concrete, let us examine typical double dot
experiments~\cite{Wilhelm00,Wilhelm02}, whose $IV$ characteristics at
$T=30\textrm{mK}$ are excerpted in Fig.~\ref{fig:comp-DQD} (a), which
is measured by applying the source-drain voltage $80\mu\textrm{V}$.
The voltage $V_{(1-2)}$ controls the degeneracy of single levels; the
topmost single orbits realize two-fold degeneracy at $V_{(1-2)}\approx
3.75, 4.25, 4.75, 5.25$ etc. The honeycomb structure clearly shows
Kondo enhancement at quarter-filling (denoting regions `b' in
Fig.~\ref{fig:comp-DQD}(a)); no enhancement at half-filling.
In Fig.~\ref{fig:comp-DQD}~(b), choosing the parameters $U=0.7
\textrm{meV}$, $U'/U=0.6$, and $U/\Gamma=20$, we plot
corresponding KR-SBMT results of the nonlinear conductance $G(T,V)$ by
varying the source-drain voltage $V=0, 35, 70\,\mu\textrm{V}$. 
The result of $V=70\,\mu\textrm{V}$ reproduces well with experimental
data; Kondo enhancement is observed at quarter-filling but not at
half-filling.  

Our KR-SBMT calculations evaluate the characteristic energy at
half-filling $T^{*}(2)$ to be $16 \mu \textrm{eV}$ for these systems.
So we conclude experiments realize large bias regime suppressing the
Kondo effect at $N_{d}\approx 2$.  It also suggests that when we can
decrease the source-drain voltage, less than $16 \mu\textrm{V}$, the
Kondo-like enhancement should appear at half-filling as well as at
quarter-filling.  

\section{Conclusion}
\label{sec:conclusion}

We have investigated the role of the inter- and intra-orbital
interactions in transport through quantum dots with two-fold orbital
degeneracy, \textit{e.g.}, carbon nanotube dots, double dot systems
with capacitive interaction etc.  
By using the $SU(4)$ Anderson model with the interaction asymmetry between
orbits/dots and within the validity of the KR-SBMT, we have shown how
significantly a small amount of it can affect the profile of the
linear conductance at finite temperature, and of the nonlinear
conductance.  We have shown that the phenomena can be understood
systematically by the different dependence $T^{*}$ on interaction
asymmetry at each gate voltage.
In addition, we have compared our theoretical results with
experimental data in two typical situations: carbon nanotube dots, and
double dot systems with capacitive interaction.  We believe our
theoretical results agree very well, suggesting unexplored phenomena
of the Kondo enhancement in double dot systems.  
We have shown clearly the interaction asymmetry is essential in
explaining conductance profiles through these two-fold degenerate
systems in real experiments.

\section{Acknowledgments}

The authors appreciate K. Isozaki, W. Izumida and H. Tamura for
helpful discussion.  One of the authors (H.O.) thanks S. Okada and K. Shiraishi
for their support.  The work was partially supported by Grant-in-Aid
for Scientific Research (Grant No. 18063003 and 18500033)
from the Ministry of Education, Culture, Sports, Science and
Technology of Japan and the Core Research for Evolutional Science and
Technology from Japan Science and Technology Corporation.


\begin{thebibliography}{99}
\bibitem{footnote2} This is believed to be an artifact of the
  Gutzwiller approximation, to which the KR-SBMT reduces at $T=0$.
\end{thebibliography}

\end{document}